 \documentstyle[12pt]{article}
 
 \newcommand{\newsection}{
 \setcounter{equation}{0}
 \section}
 \newcommand{\tr}[1]{\,{\rm tr}\,#1\,}
 \newcommand{\Sp}[1]{\,{\rm Sp}\,#1\,}
 \def\e{{\,\rm e}\,}

 \def\eop{\vspace*{\fill}\pagebreak}
 \def\be{\begin{equation}}
 \def\ee{\end{equation}}
 \def\bea{\begin{eqnarray}}
 \def\eea{\end{eqnarray}}
 \def\l{\lambda}
 \def\m {\mu}
 \def\a{\alpha}
 \def\z{\zeta}

 \newcommand{\ra}{\rightarrow}
 
 \newsavebox{\doubleline}
 \savebox{\doubleline}(30,30)[bl]
{\begin{picture}(30,30)
 \put(0,0){\vector(1,0){15}}
 \put(15,0){\vector(1,0){4}}
 \put(19,0){\line(1,0){11}}
 \put(30,0){\vector(0,1){15}}
 \put(30,15){\vector(0,1){4}}
 \put(30,19){\line(0,1){11}}
 \put(30,30){\vector(-1,0){15}}
 \put(15,30){\vector(-1,0){4}}
 \put(11,30){\line(-1,0){11}}
 \put(0,30){\vector(0,-1){15}}
 \put(0,15){\vector(0,-1){4}}
 \put(0,11){\line(0,-1){11}}
 \end{picture}}
 \title{{\bf \mbox{} \\Exact Solution to the One-Plaquette Model of Induced
 QCD at Large $N$.}
 \vspace{.5cm}}
 \author{{\bf K. Zarembo}
 \date{ }
 \vspace{.5cm} \\
 {\it Steklov Mathematical Institute} \\
 {\it Vavilov st.42, GSP-1, 117966 Moscow, Russia}}
 
\begin{document}
 \maketitle
 \vspace{-9.6cm}
 \begin{flushright}
 SMI--93--02 \\ February,   1993
 \end{flushright}
 \vspace{6.8cm}
 \begin{abstract}
 The induced lattice  gauge theory  with various types of inducing fields
 in fundamental representation of $SU(N_{c})$ is considered.
 In  a simple case of one-plaquette lattice the model  is solved
 in the  large $N_{c}$ limit by means of loop equations.
 Comparison  with the solution of usual QCD shows the  equivalence
 of induced  and  Wilson QCD providing that a mass and a number of
 flavours  of inducing fields are sufficiently large. The possibility
 to take an asymptotically free continuum limit of induced QCD  is discussed.
 \end{abstract}

 \eop
 \newsection{Introduction.}
 In this paper  we study some simlified examples of lattice gauge theories
 induced by bosons or  fermions in the fundamental representation of
 $SU(N_{c})$. The model with a single scalar field  in the adjoint
 representation received recently much attention due to the work
 \cite{KM}  where it was shown that an effective theory for inducing
 fields  is  solvable in the large $N_{c}$ limit. The further investigation
 of this model showed that it possess  a number of interesting properties
 \cite{MKA}, but the induction of a physical QCD is problematic
 \cite{KSW,KhMGM}. Models with fundamental scalars and fermions
 are also potentially solvable \cite{KSMNP,Ar}. On the other hand,
 they do not suffer from the spurious $Z_{N_{c}}$ symmetry characteristic
 for a models with adjoint inducing fields \cite{KSW}.
 The induced action obtained after integration over matter fields contains
 the traces of products of link variables along an arbitrary contours.
 However,  in the limit of large mass  of inducing fields contribution
  from a long loops is suppressed. A simple mechanism to  make the
 one--plaquette term finite in this limit was proposed by Bander
 \cite{Ban}  and Hamber \cite{Ham} and applied recently to the
 models with adjoint matter \cite{KhM}. The idea is to take
 sufficiently large number of inducing fields. The arguments of Ref.\cite
{Ban,Ham} are, however, based on the analysis  of  induced action.
 Of course, the connection between induced models and the  usual QCD
 requires a  more detailed  investigation.  In particular, it is
 desirable to have a simple  example  for which both  Wilson  and
 induced QCD can be solved exactly. Such example is provided by  a
 one--plaquette model. It  is  known that the usual QCD defined on
 a two--dimensional infinite lattice reduces to a one--plaquette
 model in  an appropriate gauge  \cite{GW}.  This  model is exactly
 solvable in the  large $N_{c}$ limit by saddle point methods \cite{GW}.
 Some other one--plaquette models were studied  in Ref.\cite{LSS,CT,AA}.
 Gross--Witten solution to the two--dimensional QCD can be reproduced
 by means of loop equations  \cite{PRF}. We use the similar approach.
 Loop equations for Kazakov--Migdal model \cite{Mak} and for some
 it's generalizations \cite{KhM} are more complicated than that for the
 model under consideration due to $Z_{N_{c}}$ symmetry. However,  in some
 degenerate cases they are equivalent. As for the one--plaquette model,
 loop equations simplifies in this case and we solve them exactly  for
 the models induced by bosons  and fermions (both chiral and Wilson).
 The comparison  with Gross--Witten solution  shows that the method of
 Ref.\cite{Ban,Ham} works nicely in  all the cases under consideration.
 The solution \cite{GW,PRF} of a two--dimensional  QCD  may be
 obtained as an  appropriate limit of the solution of induced theory.
 \newsection{Formulation of the Model and Loop Equations.}
 We shall consider the  following lattice gauge model:
 \bea

Z&=&\int\prod_{x}\prod_{f}d\varphi_{f}^{*}(x)d\varphi_{f}(x)\prod_{\m}DU_{\m}(x)
  \,\exp\left\{-N_{c}\sum_{x}\sum_{f=1}^{N_{f}}\left[m^{2}\varphi_{f}^{*}(x)
  \varphi_{f}(x)\right.\right.
  \nonumber\\
&-&\left.\left.\frac{1}{2}\sum_{\m=-D}^{D}\left(\varphi_{f}^{*}(x)

U_{\m}(x)\varphi_{f}(x+\m)+\varphi_{f}^{*}(x+\m)U_{\m}^{\dagger}(x)\varphi_{f}(x)
  \right)\right]\right\}.
 \label{2.1}
 \eea
 The $N_{f}$ scalar fields take values in the fundamental representation
 of $SU(N_{c})$ and are treated as an auxiliary fields. One should integrate
 them out to obtain an effective action for the gauge field:
 \be
 S_{ind}=\z N_{c}\sum_{C}\frac{\tr{U(C)}}{m^{2\left|C\right|}},
 \label{2.2}
 \ee
 where $\z =N_{f}/N_{c}$. Sum in (\ref{2.2}) is going over all closed contours,
 $\left|C\right|$ being a length of the contour (the number of links) and
 $U(C)$ is a path--ordered product of link variables $U_{\m}(x)$. In the
 limit \cite{Ban}
 \be
 m^{2}\ra\infty,~~~\z\ra\infty,~~~\l\equiv\frac{m^{8}}{\z}=fixed
 \label{2.3}
 \ee
 only one--plaquette terms survive with an effective gauge coupling
 $g^{2}=\l/N_{c}$.
 On the other hand, the effective theory for inducing fields obtained after
 integration over the link variables is potentially solvable in the large
 $N_{c}$ limit. In particular, the model with $N_{f}=N_{c}$ was considered
 in Ref.\cite{Ar}. In this case the inducing fields may be rearranged in a
 single complex matrix field. This permits to use a well elaborated saddle
 point techniques after the large $N_{c}$ reduction. It is worth mentioning
 that the one link integrals can be performed in the case $N_{f}\neq N_{c}$
 too \cite{BNBG}, so one may hope that the effective theory for inducing
 fields is solvable at large $N_{c}$ for an arbitrary $N_{f}$. For example,
 the model with a single ($N_{f}=1$) chiral fermion was solved in
Ref.\cite{KSMNP}.
 The loop equations (see \cite{Mig} for a review) are the large $N_{c}$ limit
 of Schwiger--Dyson identities for the Wilson loop average
 \be
 W(C)=\langle\frac{1}{N_{c}}\tr{U(C)}\rangle.
 \label{2.4}
 \ee
 To make the loop equations closed it is necessary to introduce a new gauge
 invariant quantity depending on an open path:
 \be
 G(\Gamma_{xy})=\langle\frac{1}{N_{f}}\sum_{f=1}^{N_{f}}\varphi_{f}^{*}(x)
 U(\Gamma_{xy})\varphi_{f}(y)\rangle.
 \label{2.5}
 \ee
 Performing an infinitesimal shifts $U_{\m}(x)\ra U_{\m}(x)+i\varepsilon
 U_{\m}(x)$, $\varepsilon^{\dagger}=\varepsilon$, $\varphi_{f}^{*}(x)\ra
 \varphi_{f}^{*}(x)+\eta^{*}$, $\varphi_{f}(x)\ra\varphi_{f}(x)+\eta$ and
 using the large $N_{c}$ factorization property one obtains the following
 equations:
 \bea
 \z\left[G(l_{x+\m,x}C_{xx})-G(C_{xx}l_{x,x+\m})\right]=\sum_{l\in C_{xx}}
 \tau_{\m}(l)\delta_{xy}W(C_{xy})W(C_{yx})\nonumber\\
 m^{2}G(\Gamma_{xy})-\sum_{\m=-D}^{D}G(l_{x+\m,x}\Gamma_{xy})=\delta_{xy}
 W(\Gamma_{xy})\label{loop}\\
 m^{2}G(\Gamma_{xy})-\sum_{\m=-D}^{D}G(\Gamma_{xy}l_{y,y+\m})=\delta_{xy}
 W(\Gamma_{xy}).\nonumber
 \eea
 There $l_{xy}$ stands for a link connecting $x$ with $y$, $\tau_{\m}(l)$ is
 a projection of link $l$ on the axis $\m$, it may be equal to $0$, $-1$
 or $1$. The point $y$ is defined as the beginning of $l$ if it's direction is
 positive or as the end, otherwise. The sum on the r.h.s. of the first equation
 goes over all dividings of the loop $C_{xx}$ into two parts at the point of
 selfintersection.
 One can easily verify that the following representation for the path
 amplitude is valid:
 \be
 G(\Gamma_{xy})=\frac{1}{m^{2}}\sum_{\Gamma'_{yx}}\frac{W(\Gamma_{xy}
 \Gamma'_{yx})}{m^{2|\Gamma'_{yx}|}}.
 \label{2.7}
 \ee
 The sum goes over all paths connecting $y$ with $x$. The second and the third
 equations in (\ref{loop}) are satisfied identically by this ansatz.
 The equality obtained by substitution of (\ref{2.7}) into the first equation
 is nothing that the loop equation for a model with the induced action
(\ref{2.2}).
 In the limit (\ref{2.3}) it reduces to the usual Makeenko--Migdal equation.
 Note that (\ref{2.7}) is not just the large mass expansion because Wilson
loops
 also depend on $m^{2}$. Formula (\ref{2.7}) gives the possibility to extract
 some loop averages from the correlators of inducing fields. For example,
 expanding $\langle1/N_{f}\sum_{f=1}^{N_{f}}\varphi_{f}^{*}(0)\varphi_{f}(0)
 \rangle$ in the powers of $1/m^{2}$ one can calculate the plaquette
 average in induced theory.
 The scalar fields may be replaced by fermionic ones:
 \bea

Z&=&\int\prod_{x}\prod_{f}d\overline{\psi}_{f}(x)d\psi_{f}(x)\prod_{\m}DU_{\m}(x)
  \,\exp\left\{-N_{c}\sum_{x}\sum_{f=1}^{N_{f}}\left[m\overline{\psi}_{f}(x)
  \psi_{f}(x)\right.\right.\nonumber\\

&-&\left.\left.\frac{1}{2}\sum_{\m=-D}^{D}\left(\overline{\psi}_{f}(x)P_{\m}^{-}

U_{\m}(x)\psi_{f}(x+\m)+\overline{\psi}_{f}(x+\m)P_{\m}^{+}U_{\m}^{\dagger}(x)\psi_{f}(x)
  \right)\right]\right\},
 \label{2.8}
 \eea
 where $P_{\m}^{\pm}=r\pm\gamma_{\m}$, $\gamma_{\m}$ being the Euclidean
$\gamma$
 matrices and $\gamma_{-\m}=-\gamma_{\m}$, by convention. $r=0$ corresponds to
 a chiral and $r=1$ to Wilson fermions. The induced action reads
 \be
 S_{ind}=-\z N_{c}\sum_{C}\frac{\tr{U(C)}}{m^{\left|C\right|}}\Sp{P^{-}(C)}.
 \label{2.9}
 \ee
 By Sp we denote the contraction of spinor indices and $P^{\pm}(C)$ is the
 path--ordered product of $P_{\m}^{\pm}$'s. The usual QCD is reproduced now
 in the limit \cite{Ham}
 \be
 m\ra\infty,~~~\z\ra\infty,~~~\l\equiv\frac{m^{4}}{2^{[D/2]}(1+2r^{2}-r^{4})
 \z}=fixed.
 \label{2.10}
 \ee
 Path--dependent amplitude becomes a matrix with two spinor indices:
 \be

G^{\a\beta}(\Gamma_{xy})=\langle\frac{1}{N_{f}}\sum_{f=1}^{N_{f}}\overline{\psi}_{f}^{\beta}(x)
 U(\Gamma_{xy})\psi_{f}^{\a}(y)\rangle.
 \label{2.11}
 \ee
 Loop equations for the fermionic model reads as follows:
 \bea
 \z\Sp{\left[{\bf G}(l_{x+\m,x}C_{xx})P_{\m}^{+}-P_{\m}^{-}{\bf
G}(C_{xx}l_{x,x+\m})\right]}=
 \sum_{l\in C_{xx}}
 \tau_{\m}(l)\delta_{xy}W(C_{xy})W(C_{yx})\nonumber\\
 m{\bf G}(\Gamma_{xy})-\sum_{\m=-D}^{D}{\bf
G}(l_{x+\m,x}\Gamma_{xy})P_{\m}^{+}=-\delta_{xy}
 W(\Gamma_{xy}){\bf 1}\label{2.12}\\
 m{\bf G}(\Gamma_{xy})-\sum_{\m=-D}^{D}P_{\m}^{-}{\bf
G}(\Gamma_{xy}l_{y,y+\m})=-\delta_{xy}
 W(\Gamma_{xy}){\bf 1}.\nonumber
 \eea
The analog of (\ref{2.7}) now contains the ordered product of $P_{\m}^{-}$'s:
 \be
 {\bf G}(\Gamma_{xy})=-\frac{1}{m}\sum_{\Gamma'_{yx}}\frac{W(\Gamma_{xy}
 \Gamma'_{yx})}{m^{|\Gamma'_{yx}|}}P^{-}(\Gamma'_{yx}).
 \label{2.13}
 \ee
 When the number of flavours is finite ($\z=0$) Wilson loop averages vanish
 except for the contours of zero minimal area as in the strong coupling phase
 of the models with adjoint inducing fields. Moreover, loop equations are
 equivalent in this case for the both representations of matter fields and are
 exactly solvable in any dimension \cite{Mak,KhM}.
 \newsection{The One--Plaquette Model with Scalars as an Inducing Fields.}
 In this section we consider the model defined on a lattice consisting of a
 single plaquette. It is exactly solvable in the large $N_{c}$ limit and may
 serve as a test for a more complicated models. There are five types of loop
 and path amplitudes in the model under consideration: $W_{n},S_{n},F_{n},
 G_{n}$ and $Q_{n}$ (see fig.1), where $n$ is a number of times the contour
 encircles the plaquette. Of course, these amplitudes do not depend on the
 orientation (i.e. on $\m$ and $\nu$). Loop equations (\ref{loop}) reads as
 follows:
 \bea
 \z(Q_{n-1}-F_{n})=\sum_{k=1}^{n}W_{k}W_{n-k}
&m^{2}F_{n}-G_{n}-S_{n}=0\nonumber\\
 m^{2}S_{n}-F_{n}-Q_{n-1}=W_{n}                  &m^{2}G_{n}-Q_{n}-F_{n}=0 \\
 m^{2}S_{0}-2F_{0}=1
&m^{2}Q_{n}-S_{n+1}-G_{n}=0.\nonumber
 \label{3.1}
 \eea
 After elimination of $F_{n},G_{n}$ and $Q_{n}$ one obtains the equations on
 $W_{n}$ and $S_{n}$:
 \bea
 \z(S_{n-1}-S_{n+1})=m^{2}(m^{4}-2)\sum_{k=1}^{n}W_{k}W_{n-k}\nonumber\\
 (m^{8}-4m^{4}+2)S_{n}-S_{n+1}-S_{n-1}=m^{2}(m^{4}-2)W_{n}\label{eq}\\
 (m^{8}-4m^{4}+2)S_{0}-2S_{1}=m^{2}(m^{4}-2),\nonumber
 \eea
 which may be solved by introduction of a generating functions:
 \be
 W(t)=\sum_{n=1}^{\infty}W_{n}t^{n},~~~S(t)=\sum_{n=0}^{\infty}S_{n}t^{n}.
 \label{3.3}
 \ee
 For $S(t)$ we have from (\ref{eq}):
 \be
 S(t)=\frac{m^{2}(m^{4}-2)tW(t)+\frac{1}{2}\left[(m^{8}-4m^{4}+2)S_{0}+
 m^{2}(m^{4}-2)\right]t-S_{0}}{(m^{8}-4m^{4}+2)t-1-t^{2}}.
 \label{3.4}
 \ee
 Substituting this result in the first equation in (\ref{eq}) one obtains the
 quadratic equation on $W(t)$ with a solution:
 \be
 W(t)=\frac{1}{2}\frac{(\z+1)t^{2}-(m^{8}-4m^{4}+2)t-(\z-1)+(\z-1)
 \sqrt{(t^{2}+at+1)^{2}-bt^{2}}}{(m^{8}-4m^{4}+2)t-1-t^{2}},
 \label{3.5}
 \ee
 where
 \bea
 a=\frac{(2\z-1)(m^{8}-4m^{4}+2)-\z m^{2}(m^{4}-2)(m^{4}-4)S_{0}}{(\z-1)^{2}}
 \nonumber\\
 b=a^{2}+2(m^{8}-4m^{4}+2)a-\frac{2\z-1}{(\z-1)^{2}}m^{4}(m^{4}-2)^{2}
 (m^{4}-4)+4.\nonumber
 \eea
 Loop equations do not fix $S_{0}$, it should be determined from consideration
 of an analitical structure of $W(t)$. Since $W_{n}$ are an expectation values
 of normalized traces of unitary matrices they are restricted by the condition
 $|W_{n}|\leq1$ implying that $W(t)$ is a holomorfic function of $t$ in the
unit
 circle. $W(t)$ potentially has two poles and four square root branch points.
 When one of the latter singularities lies in the unit circle the only way to
 avoid it is to put $b=0$. This corresponds to
 \be
 S_{0}^{s}=\frac{\z(m^{8}-4m^{4}+2)}{m^{2}(m^{4}-2)(m^{4}-4)}-
 \frac{\z-1}{\sqrt{m^{4}-4}}.
 \label{3.7}
 \ee
 In this case $W(t)$ has no square root singularities at all:
 \be
 W^{s}(t)=\frac{\z t}{\m-t},~~~W_{n}^{s}=\frac{\z}{\m^{n}},
 \label{3.8}
 \ee
 where
 \be
 \m=\frac{1}{2}\left[m^{8}-4m^{4}+2+m^{2}(m^{4}-2)\sqrt{m^{4}-4}\right].
 \label{3.9}
 \ee
 This is a strong coupling solution, it solves (\ref{eq}) for an arbitrary
 $\z$ and $m^{2}>2$. However, it becomes invalid when all the square root
 branch points lie on a unit circle.
 In the general case the solution (\ref{3.5}) has two branch cuts serving as a
 support of master field eigenvalue density. It is doubtful that the solution
 with a gap in the eigenvalue density should be stable, so it seems natural to
 demand the solution to have only one cut (see \cite{AA} for a rigorous proof
 in the case of Wilson fermions). This requirment is equivalent to the
condition
 $a+\sqrt{b}=2$ which fixes unambigiously $S_{0}$ to be
 \be
 S_{0}^{w}=\frac{2\z-1}{2\z}\frac{m^{4}-2}{m^{2}(m^{4}-4)}
 \label{3.10}
 \ee
 and
 \be
 W^{w}(t)=\frac{1}{2}\frac{(\z+1)t^{2}-(m^{8}-4m^{4}+2)t-(\z-1)+(\z-1)(t+1)
 \sqrt{t^{2}+2t\cos\a_{c}+1}}{(m^{8}-4m^{4}+2)t-1-t^{2}},
 \label{3.11}
 \ee
 where
 \be
 \cos\a_{c}=1-\frac{2\z-1}{2(\z-1)^{2}}m^{4}(m^{4}-4).
 \label{3.12}
 \ee
 Branch of the square root in (\ref{3.11}) is chosen so that $W^{w}(0)=0$.
 This is a weak coupling solution, it is valid when $0<a<2$ and $\z>1$. The
 latter condition is necessary because the poles at $\m$ and $1/\m$ in
 (\ref{3.11}) cancel only if $\z>1$. The critical line corresponding to the
 transition into the strong coupling phase is given by
 \be
 m_{c}^{4}(\z)=2+\frac{2\z}{\sqrt{2\z-1}}.
 \label{3.13}
 \ee
 Loop amplitudes in the weak coupling phase admit the following representation:
 \be
 W_{n}^{w}=\int_{-\a_{c}}^{\a_{c}}d\a\,\rho(\a)\cos n\a,~~~
 \rho(\a)=\frac{\z-1}{\pi}\frac{\sqrt{(\cos\a+1)(\cos\a-\cos\a_{c})}}
 {m^{8}-4m^{4}+2-2\cos\a}.
 \label{3.14}
 \ee
 One may establish it integrating the function $W^{w}(t)/2\pi it^{n+1}$ over
 the contour encircling the branch cut of $W^{w}(t)$ anticlockwise in the
 complex $t$ plane. This formula shows that $\rho(\a)$ is the density of
 eigenvalues, $\e^{i\a_{k}}$, of a vacuum--dominated configuration of gauge
 field (the master field), i.e.
$\rho(\a)=1/N_{c}\sum_{k=1}^{N_{c}}\delta(\a-\a_{k})$.
 A phase diagram of the model is devided into three regions. The region with
 $m^{2}<2$ corresponds to a Higgs phase which is unstable for the pure
quadratic
 potential. The phase transition at $m^{2}=2$ is associated with a continuum
 limit. When $\z>1$ and $2<m^{2}<m_{c}^{2}(\z)$ the weak coupling solution
 is valid, $m_{c}^{2}(\z)$ is given by (\ref{3.13}). The other values of
$m^{2}$
 and $\z$ correspond to a strong coupling phase. The phase transition between
 strong and weak coupling phases is the analog of Gross--Witten phase
transition
 in two--dimentional QCD \cite{GW}. Comparing the plaquette averages in the
both
 phases which are given by (\ref{3.8}) for the strong coupling phase and by
 \be
 W_{1}^{w}=1-\frac{m^{4}(m^{4}-4)}{4(\z-1)}
 \label{3.15}
 \ee
 for the weak coupling one, one obtains that this is a third order phase
 transition, i.e. $W_{1}$ and it's first derivatives are continious at the
 critical line (\ref{3.13}).
 At $\z=0$ there is no correlations between link variables, so this case
 corresponds to a one--dimensional model. The solution given by (\ref{3.4}),
(\ref{3.7}),(\ref{3.8}) coincides with that of Ref.\cite{Mak} at $D=1$.
 On the contrary, in the limit (\ref{2.3}) the model above describes
 two--dimensional QCD. It may be established by comparison of the solution
 (\ref{3.8}),(\ref{3.14}),(\ref{3.15}) with a results of Ref.\cite{GW,PRF}.
\newsection{The One--Plaquette Model with Fermions.}
 This model is a bit more complicated than that considered in Sec.3 because the
 path amplitudes become $2\times2$ matrices and all of them except ${\bf
S}_{0}$
 gain a dependence on $\m$ and $\nu$ (see fig.1). Eqs.(\ref{2.12}) reads as
 follows:
 \bea
 \z\Sp{({\bf Q}_{n-1}^{\nu\,\,-\m}P_{\m}^{+}-P_{\m}^{-}{\bf F}_{n}^{\m\nu})}=
 \sum_{k=1}^{n}W_{k}W_{n-k} &\nonumber\\
 m{\bf S}_{n}^{\m\nu}-P_{\m}^{-}{\bf F}_{n}^{\m\nu}-
 P_{\nu}^{-}{\bf Q}_{n-1}^{\m\nu}=-W_{n}{\bf 1}   &
 m{\bf S}_{n}^{\m\nu}-{\bf Q}_{n-1}^{\nu\,\,-\m}P_{\m}^{+}-
 {\bf F}_{n}^{-\nu\,\m}P_{\nu}^{+}=-W_{n}{\bf 1}
 \nonumber\\
 m{\bf S}_{0}-P_{\m}^{-}{\bf F}_{0}^{\m\nu}-
 P_{\nu}^{-}{\bf F}_{0}^{\nu\m}=-{\bf 1}           &
 m{\bf S}_{0}-{\bf F}_{0}^{-\m\,\nu}P_{\m}^{+}-
 {\bf F}_{0}^{-\nu\,\,\m}P_{\nu}^{+}=-{\bf 1}
 \nonumber\\
 m{\bf F}_{n}^{\m\nu}-P_{\nu}^{-}{\bf G}_{n}^{\m\nu}-
 P_{\m}^{+}{\bf S}_{n}^{\m\nu}=0                   &
 m{\bf F}_{n}^{\m\nu}-{\bf S}_{n}^{\nu\,\,-\m}P_{\m}^{+}-
 {\bf G}_{n}^{-\nu\,\m}P_{\nu}^{+}=0\\
 m{\bf G}_{n}^{\m\nu}-P_{\m}^{+}{\bf Q}_{n}^{\m\nu}-
 P_{\nu}^{+}{\bf F}_{n}^{\m\nu}=0                   &
 m{\bf G}_{n}^{\m\nu}-{\bf F}_{n}^{\nu\,\,-\m}P_{\m}^{+}-
 {\bf Q}_{n}^{-\nu\,\m}P_{\nu}^{+}=0
 \nonumber\\
 m{\bf Q}_{n}^{\m\nu}-P_{\nu}^{+}{\bf S}_{n+1}^{\m\nu}-
 P_{\m}^{-}{\bf G}_{n}^{\m\nu}=0                   &
 m{\bf Q}_{n}^{\m\nu}-{\bf G}_{n}^{\nu\,\,-\m}P_{\m}^{+}-
 {\bf S}_{n+1}^{-\nu\,\m}P_{\nu}^{+}=0.
 \nonumber
 \label{4.1}
 \eea
 Eliminating ${\bf F}_{n}^{\m\nu},{\bf G}_{n}^{\m\nu}$ and ${\bf
Q}_{n}^{\m\nu}$
 from these equations one obtains:
 \bea
 \z\Sp{}\left[\Gamma_{\m\nu}^{-}{\bf S}_{n+1}^{\m\nu}-\Gamma_{\nu\,\,-\m}^{+}
 {\bf S}_{n-1}^{\nu\,\,-\m}+\sigma(m^{2}+\sigma)({\bf S}_{n}^{\m\nu}-
 {\bf S}_{n}^{\nu\,\,-\m})\right]=m(m^{2}+2\sigma)\sum_{k=1}^{n}W_{k}W_{n-k}
 \nonumber\\
 (m^{4}+4\sigma m^{2}+2\sigma^{2}){\bf S}_{n}^{\m\nu}+
 \Gamma_{\m\nu}^{-}{\bf S}_{n+1}^{\m\nu}+
 \Gamma_{\m\nu}^{+}{\bf S}_{n-1}^{\m\nu}=-m(m^{2}+2\sigma)W_{n}{\bf 1}
 \nonumber\\
 (m^{4}+4\sigma m^{2}+2\sigma^{2}){\bf S}_{0}+
 \Gamma_{\m\nu}^{-}{\bf S}_{1}^{\m\nu}+
 \Gamma_{\m\nu}^{+}{\bf S}_{1}^{\nu\m}=-m(m^{2}+2\sigma){\bf 1}
 \label{4.2}\\
 (m^{4}+4\sigma m^{2}+2\sigma^{2}){\bf S}_{n}^{\m\nu}+
 {\bf S}_{n+1}^{\m\nu}\Gamma_{\m\nu}^{-}+
 {\bf S}_{n-1}^{\m\nu}\Gamma_{\m\nu}^{+}=-m(m^{2}+2\sigma)W_{n}{\bf 1}
 \nonumber\\
 (m^{4}+4\sigma m^{2}+2\sigma^{2}){\bf S}_{0}+
 {\bf S}_{1}^{\m\nu}\Gamma_{\m\nu}^{-}+
 {\bf S}_{1}^{\nu\m}\Gamma_{\m\nu}^{+}=-m(m^{2}+2\sigma){\bf 1},
 \nonumber
 \eea
 where $\sigma=1-r^{2}$, $\Gamma_{\m\nu}^{-}=-P_{\mu}^{-}P_{\nu}^{-}
 P_{\mu}^{+}P_{\nu}^{+}$ and $\Gamma_{\m\nu}^{+}=\Gamma_{\nu\m}^{-}$.
 Let us consider the case of chiral fermions first.
 \subsection{Chiral Fermions.}
 In this case $\Gamma_{\m\nu}^{\pm}={\bf 1}$, so ${\bf S}_{n}^{\m\nu}$
 does not depend on $\m$ and $\nu$ and are proportional to the unit matrix.
 All  the steps in the solution of eqs.(\ref{4.2}) are nearly the same as in
 the bosonic  case, so we shall not dwell on details. Introducing the
 generating functions:
 \be
 W(t)=\sum_{n=1}^{\infty}W_{n}t^{n},~~~S(t)=\sum_{n=0}^{\infty}S_{n}t^{n},~~~
 S_{n}=\Sp{{\bf  S}_{n}},
 \label{4.3}
 \ee
 we have from the eqs.(\ref{4.2}):
 \bea
 S(t)=\frac{\left[\frac{1}{2}(m^{4}+4m^{2}+2)S_{0}-m(m^{2}+2)\right]t
 +S_{0}-2m(m^{2}+2)tW(t)}{(m^{4}+4m^{2}+2)t+1+t^{2}},
 \label{4.4}\\
 W(t)=\frac{1}{2}\frac{(2\z-1)t^{2}-(m^{4}+4m^{2}+2)t-(2\z+1)+(2\z+1)
 \sqrt{(t^{2}+at+1)^{2}-bt^{2}}}{(m^{4}+4m^{2}+2)t+1+t^{2}},
 \label{4.5}
 \eea
 where
 \bea
 a=\frac{\z
m(m^{2}+2)(m^{2}+4)S_{0}+(4\z+1)(m^{4}+4m^{2}+2)}{(2\z+1)^{2}}\nonumber\\

b=a^{2}-2(m^{4}+4m^{2}+2)a+\frac{4\z+1}{(2\z+1)^{2}}m^{2}(m^{2}+2)^{2}(m^{2}+4)+4.
 \nonumber
 \eea
 The strong coupling solution reads:
 \bea
 S_{0}^{s}=-\frac{2(2\z+1)}{\sqrt{m^{2}+4}}+\frac{4\z(m^{4}+4m^{2}+2)}
 {m(m^{2}+2)(m^{2}+4)},\label{4.7}\\
 W^{s}(t)=\frac{2\z t}{\m+t},~~~W_{n}^{s}=(-1)^{n+1}\frac{2\z}{\m^{n}},
 \label{4.8}
 \eea
 where
 \be
 \m=\frac{1}{2}\left[m^{4}+4m^{2}+2+m(m^{2}+2)\sqrt{m^{2}+4}\right].
 \label{4.9}
 \ee
 At the weak coupling we have:
 \bea
 S_{0}^{w}=-\frac{(4\z+1)m}{2\z(m^{2}+2)},\label{4.10}\\
 \rho(\a)=\frac{2\z+1}{\pi}\frac{\sqrt{(\cos\a+1)(\cos\a-\cos\a_{c})}}
 {m^{4}+4m^{2}+2+2\cos\a},\label{4.11}\\
 \cos\a_{c}=1-\frac{4\z+1}{2(2\z+1)^{2}}(m^{2}+2)^{2},\label{4.12}\\
 W_{1}^{w}=1-\frac{(m^{2}+2)^{2}}{4(2\z+1)}.\label{4.13}
 \eea
 Of course, there is no Higgs phase in the fermionic model and the weak
coupling
 phase exists for all $\z>0$. The critical line is determined by
 \be
 m_{c}^{2}(\z)=\frac{2(2\z+1)}{\sqrt{4\z+1}}-2.
 \label{4.14}
 \ee
 As in the bosonic case this phase transition is of the third order. The
 $\z=0$ solution coincides with that of Ref.\cite{KSMNP,KhM} at $D=1$,
 while in the limit (\ref{2.10}) $D=2$ QCD is reproduced.
 \subsection{Wilson Fermions.}
 The solution to this model (with Wilson term added) was obtained in
 Ref.\cite{CT}. It was corrected in Ref.\cite{AA} where the model was treated
 by saddle point methods (after integration over fermions). We shall reproduce
 this solution by means of loop equations. This permits to calculate the
 fermionic amplitudes as well.
 In the case of Wilson fermions $1/4\,\Gamma_{\m\nu}^{-}$ and
 $1/4\,\Gamma_{\m\nu}^{+}$ become an orthogonal projectors. Let us introduce
 an orthonormal basis $|+,\m\nu\rangle$, $|-,\m\nu\rangle$ in which these
 projectors are diagonal. From (\ref{4.2}) it follows that ${\bf
S}_{n}^{\m\nu}$
 diagonalizes in this basis and $S_{n}^{\pm}=\langle\pm,\m\nu|{\bf
S}_{n}^{\m\nu}
 |\pm,\m\nu\rangle$ does not depend on $\m$ and $\nu$. Of course, $S_{0}^{-}=
 S_{0}^{+}\equiv S_{0}$. Introducing the generating functions:
 \be

W(t)=\sum_{n=1}^{\infty}W_{n}t^{n},~~~S^{\pm}(t)=\sum_{n=0}^{\infty}S_{n}^{\pm}t^{n}
 \label{4.15}
 \ee
 one obtains from (\ref{4.2}), taking into account that $|\pm,\m\nu\rangle=
 |\mp,\nu\m\rangle$:
 \bea
 S^{+}(t)=\frac{m^{4}S_{0}-m^{3}W(t)}{4t+m^{4}},~~~
 S^{-}(t)=\frac{4S_{0}-m^{3}t[1+W(t)]}{m^{4}t+4},
 \label{4.16}\\
 W(t)=\frac{1}{2}\frac{(\z-1)t^{2}-\frac{m^{8}+16}{4m^{4}}t-(\z+1)

+(\z+1)\sqrt{(t^{2}+at+1)^{2}-bt^{2}}}{(t+\frac{m^{4}}{4})(t+\frac{4}{m^{4}})},
 \label{4.17}
 \eea
 where
 \bea
 a=\frac{2\z m(m^{8}-16)S_{0}+(3\z+1)m^{8}+16(\z+1)}{4(\z+1)^{2}m^{4}}
 \nonumber\\
 b=a^{2}-\frac{m^{8}+16}{2m^{4}}a+\frac{(2\z+1)(m^{8}-16)^{2}}
 {16(\z+1)^{2}m^{8}}+4.
 \nonumber
 \eea
 There are two strong coupling phases, $I$ and $II$, at $m>m_{cI}(\z)$ and
 $0<m<m_{cII}(\z)$, respectively, where the critical lines are given by
 \be
 m_{cI}^{4}(\z)=4(2\z+1),~~~m_{cII}^{4}(\z)=\frac{4}{2\z+1}.
 \label{4.19}
 \ee
 The solution in these phases reads:
 \bea
 S_{0}^{sI}=-\frac{1}{m}\left(1-\frac{16\z}{m^{8}-16}\right),~~~
 S_{0}^{sII}=-\frac{\z m^{7}}{16-m^{8}},
 \label{4.20}\\
 W^{sI}(t)=\frac{4\z
t}{m^{4}+4t},~~~W_{n}^{sI}=(-1)^{n+1}\z\left(\frac{4}{m^{4}}
 \right)^{n},\nonumber\\
 W^{sII}(t)=\frac{\z m^{4}t}{4+m^{4}t},~~~W_{n}^{sII}=
 (-1)^{n+1}\z\left(\frac{m^{4}}{4}\right)^{n}.
 \label{4.21}
 \eea
 The strong coupling phases are separated by the weak coupling one at
 $m_{cII}(\z)<m<m_{cI}(\z)$. All the phase transitions are of the third
 order, it may be established by comparison of the strong coupling solutions
 with the weak coupling one:
 \bea
 S_{0}^{w}=-\frac{(4\z+1)m^{4}-4}{4\z m(m^{4}+4)},
 \label{4.22}\\
 \rho(\a)=\frac{4(\z+1)m^{4}}{\pi}\frac{\sqrt{(\cos\a+1)(\cos\a-\cos\a_{c})}}
 {m^{8}+16+8m^{4}\cos\a},
 \label{4.23}\\
 \cos\a_{c}=1-\frac{(2\z+1)(m^{4}+4)^{2}}{8(\z+1)^{2}m^{4}},
 \label{4.24}\\
 W_{1}^{w}=1-\frac{(m^{4}+4)^{2}}{16(\z+1)m^{4}}.
 \label{4.25}
 \eea
 In the limit (\ref{2.10}) the results of \cite{GW,PRF} are exactly reproduced.
 As for $D=1$ solution of Ref.\cite{KhM} which coincides in the case of Wilson
 fermions with the leading order of expansion (\ref{2.13}) due to orthogonality
 of projectors $P_{\m}^{-}$ and $P_{\m}^{+}$, it is valid only for $m^{2}>2$ at
 $\z=0$, i.e. in the strong coupling phase $I$. In the strong coupling phase
 $II$ the solution is given by (\ref{4.16}) and (\ref{4.20}) and is not
expandable
 in the powers of $1/m$.
 \newsection{Conclusions.}
 The study of the simple one--plaquette models of induced QCD shows that they
 possess a nontrivial phase structure and really induce two--dimensional QCD in
 the limit proposed by Bander and Hamber \cite{Ban,Ham}. However, all the
 consideration was resticted by the case of lattice models without any
reference
 to their continuum limit. If it is obtained in a usual way by adjusting of the
mass
 of matter fields to some critical value at fixed $\z$ difficulties with an
 asymptotic freedom may arise because due to the finitness of a critical mass
 long loops are not suppressed in the continuum limit. Such difficulties really
 arise in the fermionic one--plaquette models. In the asymptotically free
 continuum limit $\rho(\a)\ra\delta(\a)$. On the contrary, the support of
 $\rho(\a)$ is always finite at finite $m$ and $\z$ in the fermionic case.
 Although in the bosonic model $\rho(\a)\ra\delta(\a)$ as $m^{2}\ra2$, it also
 hardly induce continuum QCD in this limit due to the spurious pole in
 (\ref{3.14}) which tends to unity as $m^{2}\ra2$.
 We suggest to take the continuum limit in a different way -- it may be
achieved
 by taking the limit $\z\ra\infty$ at some large value of mass. On the one
hand,
 induced gauge coupling goes to zero in this limit and, on the other hand, long
 loops are suppressed. In the one--plaquette model $\rho(\a)\ra\delta(\a)$ in
 this limit for all considered cases.
 I am grateful to I.Ya.Aref'eva for her interest in this work and for useful
 comments. I would like to thank L.O.Chekhov and Yu.M.Makeenko for interesting
 discussions.
 
 \begin{figure}[p]
 \caption{Loop and path amplitudes in the bosonic (fermionic) one-plaquette
 model.}
 \begin{picture}(250,150)
 \multiput(10,20)(60,0){3}{\usebox{\doubleline}}
 \put(10,20){\line(-1,0){3}}
 \put(7,20){\circle*{1}}
 \put(10,20){\line(0,-1){3}}
 \put(10,17){\vector(1,0){17}}
 \put(27,17){\line(1,0){16}}
 \put(43,17){\circle*{1}}
 \put(20,7){$F_{n}({\bf F}_{n}^{\mu\nu})$}
 \put(70,20){\line(-1,0){3}}
 \put(67,20){\circle*{1}}
 \put(70,20){\line(0,-1){3}}
 \put(70,17){\vector(1,0){17}}
 \put(87,17){\line(1,0){16}}
 \put(103,17){\vector(0,1){20}}
 \put(103,37){\line(0,1){16}}
 \put(103,53){\circle*{1}}
 \put(80,7){$G_{n}({\bf G}_{n}^{\mu\nu})$}
 \put(130,20){\line(-1,0){3}}
 \put(127,20){\circle*{1}}
 \put(130,20){\line(0,-1){3}}
 \put(130,17){\vector(1,0){17}}
 \put(147,17){\line(1,0){16}}
 \put(163,17){\vector(0,1){20}}
 \put(163,37){\line(0,1){16}}
 \put(163,53){\vector(-1,0){20}}
 \put(143,53){\line(-1,0){16}}
 \put(127,53){\circle*{1}}
 \put(140,7){$Q_{n}({\bf Q}_{n}^{\mu\nu})$}
%%%%%%%%%%%%%%%%%%%%%%%%%%%%%%%%%
 \thicklines
 \put(10,90){\vector(1,0){30}}
 \put(10,90){\vector(0,1){30}}
 \thinlines
 \put(36,86){$\mu$}
 \put(5,116){$\nu$}
 \put(70,90){\usebox{\doubleline}}
 \put(85,80){$W_{n}$}
 \put(130,90){\usebox{\doubleline}}
 \put(130,90){\line(-1,0){3}}
 \put(127,90){\circle*{1}}
 \put(130,90){\line(0,-1){3}}
 \put(130,87){\circle*{1}}
 \put(140,80){$S_{n}({\bf S}_{n}^{\mu\nu})$}
 \end{picture}
 \end{figure}
 \end{document}